\RequirePackage{fix-cm}
\documentclass{svjour3}                     
\smartqed  
\usepackage{graphicx}
\usepackage{braket}
\usepackage{graphicx}
\usepackage{subcaption}
\usepackage{amsmath}
\usepackage{amsfonts}
\usepackage{booktabs}
\usepackage{multirow}
\usepackage{makecell}
\usepackage{xurl}
\usepackage{verbatim}

\usepackage{mathptmx}      
\begin{document}

\title{RGB Image Classification with Quantum Convolutional Ansatz
}


\author{Yu Jing         \and
        Xiaogang Li     \and
        Yang Yang		\and
        Chonghang Wu	\and
        Wenbing Fu		\and
        Wei Hu			\and
	   Yuanyuan Li \and
        Hua Xu* \thanks{Corresponding author: 
        Hua Xu, hua.xu@kfquantum.com}
}


\institute{Yu Jing \and Xiaogang Li \and Yang Yang \and Chonghang Wu \and Wenbing Fu \and Wei Hu \and Hua Xu \at
              Kunfeng Quantum Technology Co., Ltd, Shanghai, China. \\
              \email{yu.jing, xiaogang.li, yang.yang, chonghang.wu, wenbing.fu, wei.hu and hua.xu@kfquantum.com} \\          
		\and
		Yuanyuan Li \at
		School of Electronic and Electrical Engineering, Shanghai University of Engineering Science, Shanghai, China \\
		\email{liyuanyuan@sues.edu.cn}
		}

\date{Received: date / Accepted: date}

\maketitle

\begin{abstract}
With the rapid growth of qubit numbers and coherence times in quantum hardware technology,
implementing shallow neural networks on the so-called Noisy Intermediate-Scale Quantum (NISQ) devices has attracted a lot of interest.
Many quantum (convolutional) circuit ansaetze are proposed for grayscale images classification tasks with promising empirical results.
However, when applying these ansaetze on RGB images, the intra-channel information that is useful for vision tasks is not extracted effectively.
In this paper, we propose two types of quantum circuit ansaetze to simulate convolution operations on RGB images, which differ in the way how inter-channel and intra-channel information are extracted.
To the best of our knowledge, this is the first work of a quantum convolutional circuit to deal with RGB images effectively, with a higher test accuracy  compared to the purely classical CNNs.
We also investigate the relationship between the size of quantum circuit ansatz and the learnability of the hybrid quantum-classical convolutional neural network.
Through experiments  based on CIFAR-10 and MNIST datasets, we demonstrate that a larger size of the quantum circuit ansatz improves predictive performance in multiclass classification tasks,
providing useful insights for near term quantum algorithm developments.

\end{abstract}

\section{Introduction} \label{intro}
Quantum processors with exponential speedup potentials have been released in recent years, e.g. IBM released IBM Q $53$  \cite{Garc2020} that contains $53$ qubits, which greatly motivates research on developing computational models that are applicable on quantum devices.
Many quantum circuits are proposed to implement quantum counterparts of
classical machine learning algorithms, such as support vector machine
\cite{Rebentrost_2014}, k-nearest neighbor \cite{wiebe2015quantum}, and k-means clustering   \cite{Lloyd2013QuantumAF}.
Among these quantum machine learning algorithms, quantum (convolutional) neural networks can be implemented with near-term quantum devices relatively easier because of their noise-tolerant characteristic and lower requirement for circuit depth.

Convolution operation is the most decisive component of multiple famous deep neural networks such as VGG \cite{simonyan2014very}, ResNet \cite{he2016deep}, and etc, to achieve human-alike performances on computer vision tasks.
Recently, the implementation of a convolution operation on quantum processors has rapidly attracted interest of researchers. Cong et al. \cite{cong2019quantum} propose the first quantum convolutional neural network that simulates 1D convolution operation with a quantum circuit. Liu et al. \cite{liu2019hybrid} extend such a circuit to simulate 2D convolution operation which is capable of processing image data. Oh et al. \cite{oh2020tutorial} train a hybrid quantum-classical convolutional neural network on the whole MNIST dataset which achieves good predictive accuracy close to what classical convolutional neural network does.  After that, designing quantum circuit ansaetze and bulidng hybrid quantum neural networks to process images has gained increasing attention. Tacchino et al. \cite{tacchino2020quantum} implement a quantum feed-forward neural network to classify $2 \times 2$ images into two categories with a 7-qubit superconducting processor. Zhao and Gao \cite{zhao2021qdnn} train a quantum neural network on a classical computer to classify two digits from the MNIST dataset which is downsized to size $8 \times 8$. Stein et al. \cite{stein2021quclassi} construct a hybrid quantum neural network with a quantum state fidelity-based cost function, and trained a binary and a 10-class classifier with the MNIST dataset. Jiang et al. \cite{jiang2020co} provide a procedure to encode pixels from images and transform trained classical neural networks to their corresponding quantum-form network. 


As quantum hardware with sufficient qubits and coherence time for meaningful real applications is still years away from materializing, researchers usually simulate and train quantum convolutional neural networks utilizing classical computers with quantum simulators, which are subject to resource constraints. A quantum state of $n$ qubits requires \(2^{3+n}\) bytes storage if simulated with the double-precision requirement, and quantum operations on $n$ qubits are unitary matrices of size  $ 2^n \times 2^n $. Due to such computation power limitations, previous researchers usually use a small number of qubits in experiments, and compressed images to a very small size while carrying out model training, e.g. from size $28 \times 28$ to size $10 \times 10$ \cite{oh2020tutorial}. Most of the available works are designed only for classifying greyscale images, and there are not any quantum convolutional neural networks designed for classifying RGB images as far as we know. In addition, the relationship between the kernel size of the convolutional quantum circuit and the classification performance of the overall neural network is an interesting question worth investigating. Yetics et al. \cite{yetics2020quantum} pioneered different sizes of quantum circuits for a binary grey-image classification problem. Pesah et al. \cite{pesah2020absence} provide a theoretical guarantee that for quantum convolutional neural network, gradients of weights vanish polynomially w.r.t the size of the quantum circuit ansatz. However, till now, the relevant work is quite limited, and how the size of the quantum convolution layer impacts its trainability still needs further investigation. 

Recently, with the general availability of cloud quantum simulation platforms such as the Amazon Braket \cite{braket2021}  and Kunfeng Kungang cloud platform \cite{kungang},  we are able to alleviate such computation resource limitation and conduct empirical simulations at a much larger scale. We design two quantum circuit ansaetze for RGB images and demonstrate that more qubits entangled in the ansaetze (larger kernel size) improve the performance of hybrid quantum-classical convolutional neural networks, which provides useful insights for near term quantum algorithm developments. The major contributions of this work are:

\begin{itemize}
	\item We propose two types of quantum circuits to mimic convolution operations, which can be applied to images with RGB channels.
	To the best of our knowledge, this is the first design of a quantum circuit dealing with colorful images effectively.
	\item We investigate and analyze the relationship between the size of  the quantum convolutional layer (number of qubits in the circuit )
	and the classification performance of the overall hybrid quantum-classical neural network.
	\item We conduct  experiments on Kunfeng's  large scale parallel simulation platform  using the CIFAR-10 and MNIST image datasets and demonstrate a higher test accuracy  compared to the purely classical CNNs.
\end{itemize}

\section{Background} \label{bkgrd}

\subsection{Quantum Convolutional Circuit} \label{quantum_conv}

\noindent \textbf{Convolution operation} For classical convolution operation applied on image processing tasks, a kernel with size $S \times T$ is defined as a feature extractor that scans a whole image $X \in \mathbb{R}^ {M \times N}$, and $x_{i, j}$ is the value of the $(i,j)^{th}$ pixel in the image $X$.
The kernel is denoted as a matrix $W \in \mathbb{R} ^{S \times T}$ which contains weights $w_{i,j}$ to be trained.
By calculating the dot product of the kernel and the image slice where the kernel covers, together with the activation function $\sigma$, the feature map from this image slice is extracted.
The convolution operation can be denoted as a mapping from $X$ to feature maps

\begin{equation}
	\begin{gathered}
		f_{i, j} := \sigma(A_{i:i+S, j: j+T} W),
	\end{gathered}
\end{equation}

\noindent in which $A_{i:i+S, j: j+T}$ is the slice of image $X$ that the kernel $W$ covers, $1 \leq i \leq M$ and $1 \leq j \leq N$.

By defining multiple kernels, each one is expected to extract features that have different characteristics.
When these multiple kernels are grouped to process the same input feature maps, a \textit{convolutional layer} is formed. Without specification, $W$ represents multiple kernels in this paper.

\noindent \textbf{Quantum Convolutional Circuit} To mimic convolution operation with a quantum circuit, there are different designs of quantum circuit ansatz as mentioned in Section \ref{intro}.
We briefly introduce the quantum convolutional circuit here\cite{cong2019quantum,liu2019hybrid,oh2020tutorial}.

The ways to construct quantum convolutional circuit ansaetze are similar, most of which are consisted of three modules.
The \textit{first module}, the quantum ansatz contains $Q$ qubits which act as a foundation to imitate a classical convolution operation.
The number of qubits in the ansatz is often referred to as the size of ansatz.
The \textit{second module} is designed to encode input for further processing with the quantum convolutional circuit. The input can be pixel values or feature maps from the previous layer.
Single-qubit rotation gates, e.g. Rx gate, are defined with the qubits in ansatz.
The input pixel values are treated as the rotation angles for these single-qubit rotation gates.
The \textit{third module} is the entanglement stage for imitating convolution operation.
In this module, the cluster of multi-qubits controlled gates with unknown parameters is designed to extract task-specific features.
Then with predefined decode function $Z$ (or measurement), usually Pauli gates, features are decoded for the subsequent layer.
More specifically, the expectation of the Pauli gates is the task-specific feature.

The quantum convolutional circuit ansatz described above can be represented mathematically as below.
Let us denote all $M$ controlled unitary operations with unknown parameters in the quantum circuit as $\mathbf{U}_1(\theta_1), \mathbf{U}_2(\theta_2), ...,$, $ \mathbf{U}_M(\theta_M)$.
Since unitary property remains during the product of unitary matrices, we use $\mathbf{U}(\theta)$ to denote all unitary operations in the circuit for simplicity.
Then quantum convolution operation can be denoted as,

\begin{equation}
	\begin{gathered}
		E = \braket{\psi | \mathbf{U}^{\dag}(\mathbf{\theta}) \mathbf{Z}^{\otimes Q} \mathbf{U}(\mathbf{\theta})| \psi},
	\end{gathered}
\end{equation}

\noindent in which $E$ represents expectation observed, $\ket{\psi}$ is the input quantum state, and $Q$ is the subset of all qubits in the ansatz.
The observation could be just for one qubit or for all qubits in the ansatz.

The subsequent layer in the quantum convolutional neural network takes the expectation $E$ as its input.
Same as the classical convolutional layer, multiple quantum circuits could be defined to extract distinct features from the same input, which forms a \textit{Quantum Convolutional Layer}.

\subsection{Hybrid Quantum-classical Neural Network}

While constructing a hybrid quantum-classical neural network, the quantum circuit ansatz can be placed at the beginning of the network or inserted as an intermediate layer in the network. Take the former case as an example, the hybrid quantum-classical convolutional neural network that contains $l$ layers can be denoted as follow:

\begin{equation}
	\begin{gathered}
		F_1= \sigma_1 (W_1 E + b_1), \\
		F_2= \sigma_2 (W_2 F_1 + b_2), \\
		\vdots \\
		\hat{Y} = \sigma_l (W_l F_{l-1} + b_l),
	\end{gathered}
\end{equation}
where $b_1 , b_2 , \dots,  b_l$ are biases and $\hat{Y}$ is the vector of predicted values.

\noindent Then the loss function can be defined as:

\begin{equation}
	\begin{gathered}
		L(X, W; \theta) = crossEntropy(Y, \hat{Y}).
	\end{gathered}
\end{equation}

\noindent Then the partial derivatives can be calculated to update unknown parameters in the hybrid quantum-classical convolutional neural network. The gradients of the unknown parameters from the quantum circuit are calculated as \cite{liu2019hybrid},

\begin{equation}
	\begin{gathered}
		\frac {\partial E} {\partial \theta_i}
		= \frac{1}{2} (\braket{ \psi | \mathbf{U}^{\dag}(\theta^{+}_i) \mathbf{Z}^{\otimes Q} \mathbf{U}(\theta^{+}_i) | \psi} \\
		- \braket{ \psi | \mathbf{U}^{\dag}(\theta^{-}_i) \mathbf{Z}^{\otimes Q} \mathbf{U}(\theta^{-}_i) | \psi}),
	\end{gathered}
\end{equation}

\noindent in which $\theta^{\pm}_i := \theta_i \pm \frac{\pi}{2}$.
By following the chain rule, $\frac {\partial L} {\partial \theta_i}$ can be calculated. With a deep learning framework such as TensorFlow or PyTorch, an end-to-end learning strategy can be applied to learn unknown parameters in the quantum circuits.

\begin{figure*}[htpb]
	\centering
	\begin{subfigure}[b]{\linewidth}
		\centering
		\includegraphics[width=0.95\linewidth]{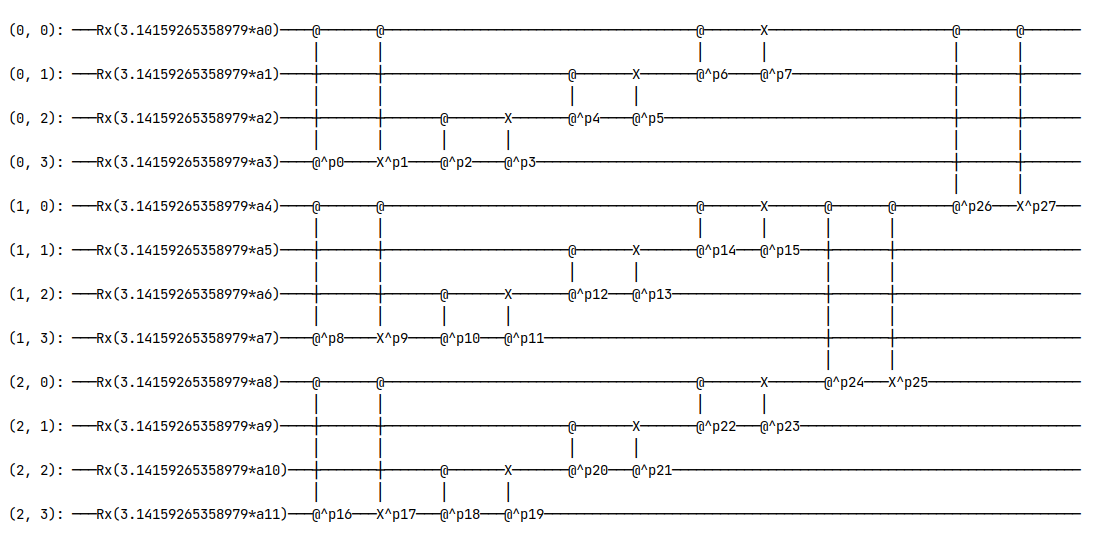}
		\caption{Quantum circuit of HQconv}
		\label{fig:HQconv}
	\end{subfigure}
	\begin{subfigure}[b]{\linewidth}
		\centering
		\includegraphics[width=0.95\linewidth]{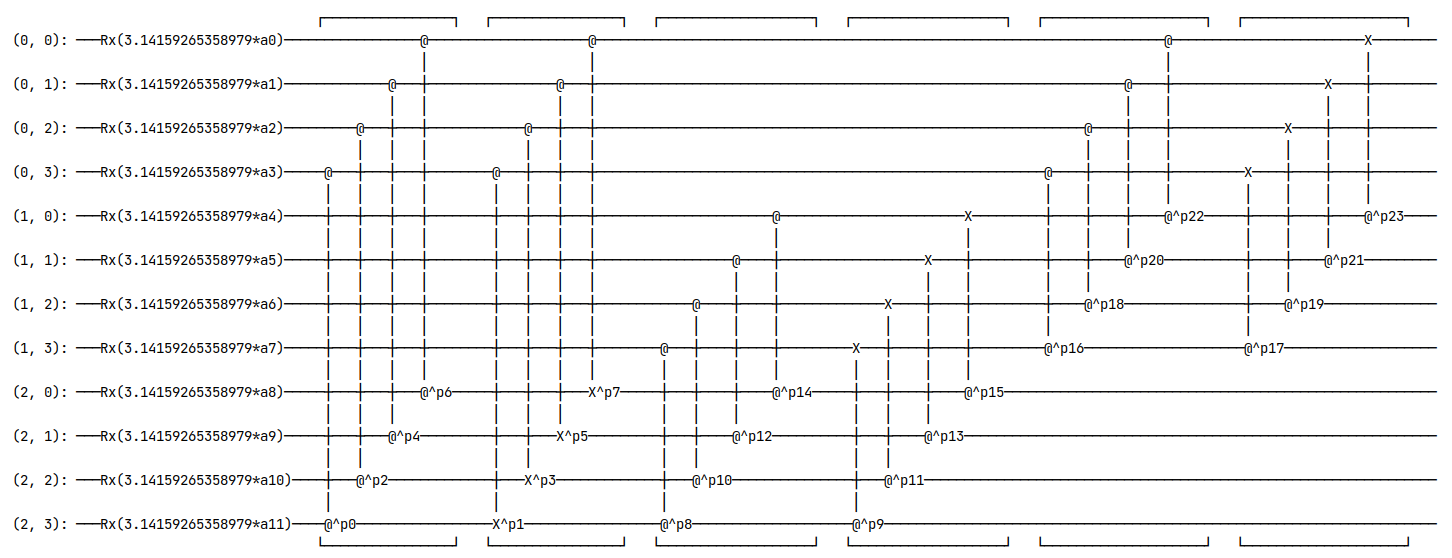}
		\caption{Quantum circuit of FQconv}
		\label{fig:FQconv}
	\end{subfigure}
	\caption{Sample circuits of HQconv and FQconv. Qubit $(i, j)$ represents qubit that corresponds to the $j^{th}$ pixel from $i^{th}$ channel. Single qubit Rx gates take pixel values as parameters, p1, p2, ..., p27 represent unknown parameters in two-qubits controlled gates. (a) HQconv, stride is $1$. (b) FQconv, stride is $4$.}
	\label{fig:FQconv vs HQconv}
\end{figure*}

\section{Methods} \label{methods}


The RGB color model is a common model used to create color digital images, which are made of pixels.
Each pixel is represented with a combination of three integer numbers that stand for three primary color channels: red, green, and blue.
By varying these three values, enormous colors can be reconstructed.
The inter-channel information discloses the closeness between two pixels on that color channel, e.g. the sharp variation among adjacent pixels normally stands for an edge or the change of object; and the intra-channel information presents the exact perceptional color of that pixel. Both are crucial for recognizing objects.

Our methods are rooted in Liu et al. \cite{liu2019hybrid}.
To the best of our knowledge, this is the first quantum ansatz designed to handle colorful images.
The quantum convolutional ansatz we propose is capable to extract features that combine inter-channel and intra-channel information.
As we find that features extracted by previous quantum ansatz intend to underperform in classification tasks without considering all three channels as a whole.
For example, using identical quantum ansatz for all three channels or three different quantum ansatzes for three individual channels is not effectively enough to encode intra-channel information.

In our methods, a few extra two-qubit controlled gates are utilized to encode intra-channel information.
With regards to the ways of extracting inter-channel or intra-channel information, we design two types of quantum convolutional ansaetze: the Hierarchical Quantum Convolutional Ansatz (HQconv) and the Flat Quantum Convolutional Ansatz (FQconv).
HQconv extracts local information separately within individual channels first. Once done, extra controlled gates are used to encode intra-channel information.
Therefore HQconv naturally has a hierarchical topology.
FQconv encodes intra-channel information and inter-channel information at the same time, which makes its topology structure flat.

Instead of using extra quantum controlled gates in HQconv, a hyperparameter called stride is employed to control how much intra- or inter-channel information is studied.
Stride represents the distance between the control qubit and the target qubit in the controlled quantum gate.
When the stride value is small, the control qubits and their corresponding target qubits are from the same image channel for most of the controlled gates in the quantum circuit.
Only for a few controlled gates, the control qubits and target qubits are from two consecutive image channels.
In this case, local inter-channel information is  examined more than the intra-channel information. Similarly,  by enlarging stride value, intra-channel information is examined more, and inter-channel information can be ignored completely when the stride is larger than the size of the kernel.
The concept of stride also can be applied to HQconv as long as its value is smaller than the total number of qubits in that channel.

As shown in Figure \ref{fig:HQconv} and Figure \ref{fig:FQconv}, which are two examples of HQconv and FQconv, single-qubit unitary rotation operations are applied to encode pixel values, and controlled unitary operations with unknown parameters are performed to encode the relationship between pixels.
In Section \ref{exprim}, we compare HQconv and FQconv w.r.t their classification performance.

\subsection{Mathematical Respresentation of HQconv}
Here we describe HQconv mathematically. Let $x_{p, c} \in A_{s: s+S, t:t+T}$ $(1 \leq p \leq S \cdot T, 1 \leq c \leq 3)$, where $A_{s: s+S, t:t+T} $ is one slice of a RGB image and $x_{p, c}$ is the value of $p^{th}$ pixel from $c^{th}$ channel. For simplicity, we omit the subscript of $A$.

\noindent Let ${q_{1, 1}, ..., q_{p, c}, ..., q_{S\cdot T, 3}}$ be qubits in HQconv ansatz and

\begin{equation} \label{eq: init}
	\begin{gathered}
		\ket{\psi^0}:= \bigotimes_{p,c} \ket{q^0_{p, c}}
	\end{gathered}
\end{equation}

\noindent be the initial state of all qubits, the superscripts of $q^0_{p, c}$ and $\psi^0$ represent the initial state of qubit $q_{p,c}$ and the state of all qubits in the ansatz.
Encoding of input pixel value $x_{p, c}$ is done by utilizing Rx gate (denoted as $\mathbf{R_x}$) defined on qubit $q_{q, c}$.
Then the state of $q_{q, c}$ after Rx gate can be represented as:

\begin{equation}
	\begin{gathered}
		\ket{q^1_{p, c}}= \mathbf{R_x}(x_{p,c} \cdot \pi)\ket{q^0_{p, c}},
	\end{gathered}
\end{equation}

\noindent in which
\begin{equation}
	\begin{gathered}
		\mathbf{R_x}(x_{p,c} \cdot \pi)=
		\begin{bmatrix}
			\cos(x_{p,c} \cdot \pi/2) & -i\sin(x_{p,c} \cdot \pi/2) \\
			-i\sin(x_{p,c} \cdot \pi/2) & \cos(x_{p,c} \cdot \pi/2)
		\end{bmatrix}.
	\end{gathered}
\end{equation}

\noindent  The states of the ansatz after processing input can be denoted as

\begin{equation} \label{eq: encode}
	\begin{gathered}
		\ket{\psi^1} := \ket{q^1_{1,1},q^1_{2,1},...,q^1_{p,c},...,q^1_{S \cdot T, 3}} \\
		= \bigotimes_{p,c} \mathbf{R_x}(x_{p,c} \cdot \pi)\ket{q^0_{p, c}}
	\end{gathered}
\end{equation}

\noindent As mentioned above,  in HQconv  two-quit controlled gates are used to extract information within each channel first,
with  an extra controlled gate to code information between channels afterwards.
Let $\mathbf{C_z}$ and $\mathbf{C_x}$ represent CZPowgate and CXPowgate, $strd$ represents for the value of hyperparameter stride,
then for any control qubit $q_{p, c}$ and target qubit $q_{p+strd, c}$ $(p+strd\leq S \cdot T)$,

\begin{equation} \label{eq: inter-C control}
	\begin{gathered}
		\ket{q^{2}_{p, c},q^{2}_{p+strd, c}} = (\mathbf{C_x}(\theta_{x, p, c}) \circ \mathbf{C_z}(\theta_{z, p, c})) \ket{q^1_{p, c},q^1_{p+strd, c}},
	\end{gathered}
\end{equation}

\noindent in which $\theta_{x, p, c}$, $\theta_{z, p, c}$ are the unknown parameter to be learnt later,
$\circ$ represents the composition of gate $\mathbf{C_x}$ and gate $\mathbf{C_z}$, $\mathbf{C_z}$ is implemented first.
Note there exists an entanglement between the qubit $q^{2}_{p, c}$ and the qubit
$q^{2}_{p+strd, c}$ at this stage, so the state  $\ket{q^{2}_{p, c},q^{2}
	_{p+strd, c}}$  in Eq. \ref{eq: inter-C control} is inseparable.

The matrix forms of $\mathbf{C_x}$ and $\mathbf{C_z}$ are,
\begin{equation}
	\begin{gathered}
		\mathbf{C_x}(\theta_{x, p, c}) =
		\begin{bmatrix}
			1 & 0 & 0 & 0 \\
			0 & 1 & 0 & 0 \\
			0 & 0 & \mathsf{g} \cdot \mathsf{c} & - i \cdot \mathsf{g} \cdot \mathsf{s}  \\
			0 & 0 & - i \cdot \mathsf{g} \cdot \mathsf{s} & \mathsf{g} \cdot \mathsf{c}
		\end{bmatrix},
	\end{gathered}
\end{equation}

\noindent and
\begin{equation}
	\begin{gathered}
		\mathbf{C_z}(\theta_{z, p, c}) =
		\begin{bmatrix}
			1 & 0 & 0 & 0 \\
			0 & 1 & 0 & 0 \\
			0 & 0 & 1 & 0\\
			0 & 0 & 0 & e^{i\pi\theta_{z, p, c}}
		\end{bmatrix}.
	\end{gathered}
\end{equation}

\noindent in which
\begin{equation}
	\begin{gathered}
		\mathsf{g} := e^{i\pi\theta_{x, p, c}}, \\
		\mathsf{c} := \cos (\pi \cdot \theta_{x, p, c} / 2),	\\
		\mathsf{s} := \sin (\pi \cdot \theta_{x, p, c} / 2).
	\end{gathered}
\end{equation}

\noindent Then the state of ansatz after encoding inter-channel information can be denoted as

\begin{equation}
	\begin{gathered}
		\ket{\psi^2} :=  \bigotimes_{p \leq \frac{S \cdot T}{2},c} \ket{q^{2}_{p,
				c},q^{2}_{p+strd, c}}.
	\end{gathered}
\end{equation}

\noindent By including extra CZPowgates and CXPowgates, intra-channel information can be processed, then the state of ansatz is

\begin{equation} \label{eq: intra-C control}
	\begin{gathered}
		\ket{q^{3}_{1, c-1},q^{3}_{1, c}} = (\mathbf{C_x}(\theta_{x, c}) \circ \mathbf{C_z}(\theta_{z, c})) \ket{q^2_{1, c-1},q^2_{1, c}},
	\end{gathered}
\end{equation}

\noindent in which $q^2_{1, c-1}$ is the control qubit and $q^2_{1, c}$ is the target qubit.
Evaluating Eq. \ref{eq: intra-C control} to all possible $c$, the state of the ansatz becomes,

\begin{equation} \label{eq: after intra-c}
	\begin{gathered}
		\ket{\psi^3} := \ket{q^3_{1, 1},q^3_{2, 1},...,q^3_{S \cdot T, 3}}.
	\end{gathered}
\end{equation}

\noindent After applying PauliZ gate, the expectation is

\begin{equation}
	\begin{gathered}
		E = \braket{\psi^3 | \mathbf{Z}^{\otimes Q} |\psi^3},
	\end{gathered}
\end{equation}

\noindent where $Q$ represents a subset of all qubits in the ansatz.

\subsection{Mathematical Respresentation of FQconv}
Let $x_p$ be the value of pixels from image slice $A$, $1 \leq p \leq S \cdot T \cdot C$.
For FQconv, since the way of encoding input pixel values is the same as that in HQconv,
the calculation routines from Eq. (\ref{eq: init}) to Eq. (\ref{eq: encode}) are the same here.
The procedure to extract features from inter- and intra-channel can be represented as:

\begin{equation} \label{eq: fqconv control}
	\begin{gathered}
		\ket{q^{2}_{p},q^{2}_{p+strd}} = (\mathbf{C_x}(\theta_{x, p}) \circ \mathbf{C_z}(\theta_{z, p})) \ket{q^1_{p},q^1_{p+strd}},
	\end{gathered}
\end{equation}

\noindent in which $q^1_{p}$ is the control qubit and $q^1_{p+strd}$ is the target qubit, $p+strd \leq S \cdot T \cdot C$.
Then the state of the ansatz after this procedure is

\begin{equation} \label{eq: fqconv after control}
	\begin{gathered}
		\ket{\psi^2} :=  \bigotimes_{p \leq \frac{S \cdot T \cdot C}{2}} \ket{q^{2}_{p},q^{2}_{p+strd}}.
	\end{gathered}
\end{equation}

\noindent By including PauliZ gate, the expectation is

\begin{equation}
	\begin{gathered}
		E = \braket{\psi^2 | \mathbf{Z}^{\otimes Q} |\psi^2},
	\end{gathered}
\end{equation}
\noindent where $Q$ represents a subset of all qubits in the ansatz.

Similarly, the states represented in Eq. \ref{eq: intra-C control}, Eq. \ref{eq:
	after intra-c}, Eq. \ref{eq: fqconv control} and Eq. \ref{eq: fqconv after control} are highly entangled and thus inseparable, like the one we mentioned in Eq. \ref{eq: inter-C control}.

\begin{figure}[htpb]
	\centering
	\includegraphics[width=0.6\linewidth]{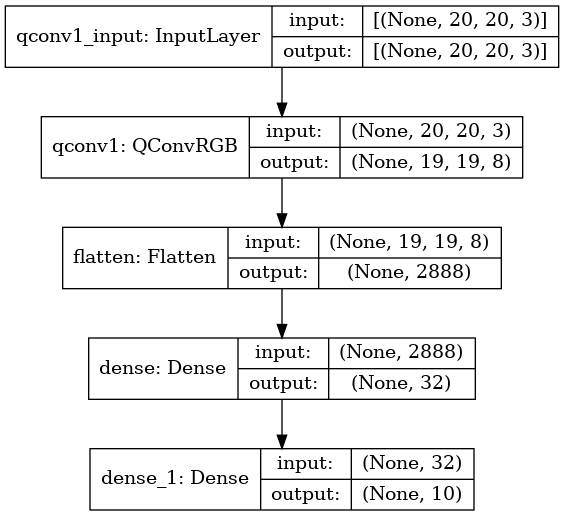}
	\caption{Architecture of the hybrid quantum-classical convolutional neural network using quantum circuit ansatz. (kernel size $2 \times 2$)}
	\label{fig:QCNN}
\end{figure}

HQconv and FQconv differ in the way  how inter-channel and intra-channel information are extracted from images.  Through experiments, we demonstrate that the order to process inter-channel and intra-channel information, as well as the encoding of intra-channel information is critical to improve the performance of the hybrid quantum-classical convolutional neural networks.

\section{Experiments} \label{exprim}

To evaluate the implementation of  our proposed quantum circuit ansaetze and the techniques presented in this paper,  two groups of experiments with distinct purposes are conducted.
In  \textit{Experimental Group A}, hybrid quantum-classical convolutional neural networks are constructed for experiments. 
Both noiseless and noisy scenarios are considered during the experiments to provide more comprehensive evidence for evaluating the effectiveness of HQconv and FQconv.
We also compare these models with the corresponding classical counterparts. 

In \textit{Experimental Group B}, with the identical structure of hybrid models, we investigate how the size of quantum convolutional ansatz (kernel size) impacts the performance of the hybrid model in multi-class classification tasks.

\subsection{Experiments Setup} \label{sub}

\begin{figure}[htpb]
	\centering
	\includegraphics[width=0.6\linewidth]{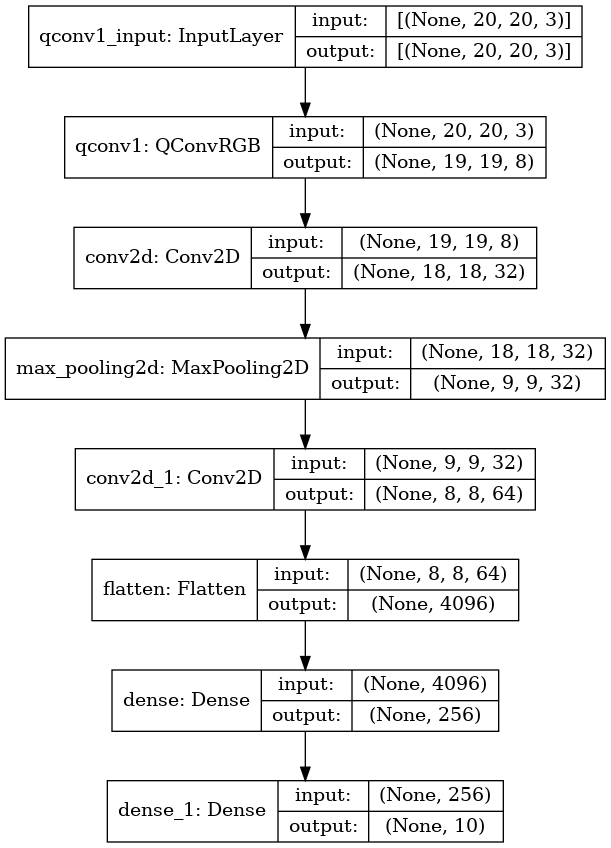}
	\caption{The other architecture of the hybrid quantum-classical convolutional neural network using quantum circuit ansatz. (kernel size $2 \times 2$) }
	\label{fig:NQCNN archit}
\end{figure}

\noindent \textbf{Datasets} Three datasets used in experiments are listed as below:

\begin{itemize}
	\item CIFAR-10-small (used in Experimental Group A \& B): an RGB image dataset that contains 50k images as training samples and 10k images as testing samples.
	For experimental purposes, 200 images are randomly sampled from all training samples to form the dataset `CIFAR-10-small'.
	Unless stated otherwise, CIFAR-10 in this paper refers to CIFAR-10-small dataset.
	For different purposes, we compress images to size $10 \times 10$ or $20 \times 20$.
	\item MNIST (used in Experimental Group B): a greyscale image dataset contains digits from `0' to `9'. There are 60k training samples and 10k testing samples.
	\item Fashion-MNIST (used in Experimental Group B): a greyscale image dataset contains 10 basic shapes of t-shirts, dresses, shoes, etc.
	There are 60k training samples and 10k testing samples.
\end{itemize}

\noindent \textbf{Architecture of hybrid quantum-classical convolutional neural network} For most of our experiments, architectures of the hybrid quantum-classical convolutional neural networks are identical (as shown in Figure \ref{fig:QCNN}).
The hybrid models contain one quantum convolutional layer with $8$ filters without specification, and two fully-connected layers with $32$ and $10$ hidden units.
Some experiments use architecture with classical convolutional layers as shown in Figure \ref{fig:NQCNN archit}.
The activation function used is ReLU for all layers except Softmax is used for the last fully-connected layer for both architectures shown in Figure \ref{fig:QCNN} and Figure \ref{fig:NQCNN archit}.

\noindent \textbf{Benchmark models} For Experimental Group A, the benchmark models are the corresponding classical convolutional neural networks that share similar architectures with the hybrid quantum-classical convolutional neural networks.
The quantum convolutional layer is replaced with a classical convolutional layer which has the same number of filters.
For example, the structure of the corresponding classical model for architecture shown in Figure \ref{fig:NQCNN archit}  is CONV1(8) - CONV2(32) - POOL1 -CONV3(64) - FC1(256) - FC2(10).
The numbers in the parentheses represent the number of filters in the convolutional layers or the number of hidden units used in the fully-connected layers.
The activation function used is ReLU for all layers except Softmax is used for the last fully-connected layer.
The kernel size for all convolutional layers and the pooling layer is $(2, 2)$.
For Experimental Group B, we focus on comparing quantum ansaetze with different
kernel sizes, so there are no benchmark models for the experiments.

\noindent \textbf{Evaluation merits} Prediction accuracy is the major evaluation merit.
Confusion matrices are also provided for some experiments.

\noindent \textbf{Hyperparameters} The choices of parameters are listed as below:
\begin{itemize}
	\item The number of filters in the quantum convolutional layer is eight by default, unless stated otherwise.
	\item Stride for Experimental Group A, HQconv uses stride $1$ or $2$, FQconv uses stride from $1$ to $6$; for Experimental Group B, HQconv with stride $1$ is used through all experiments.
\end{itemize}


\begin{figure}[htpb]
	\centering
	\begin{subfigure}[b]{0.4\linewidth}
		\includegraphics[width=\linewidth]{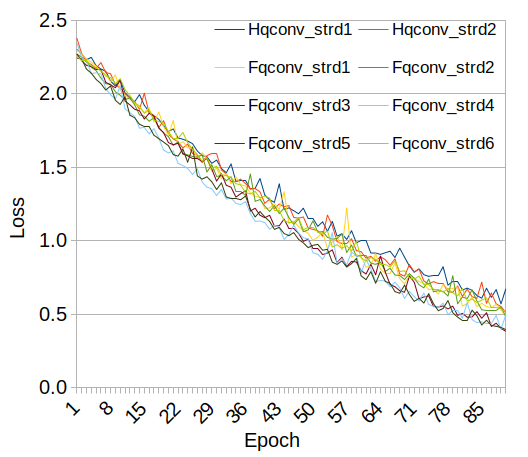}
		\caption{\centering Loss of HQconvs and FQconvs \newline}
	\end{subfigure}
	\begin{subfigure}[b]{0.4\linewidth}
		\includegraphics[width=\linewidth]{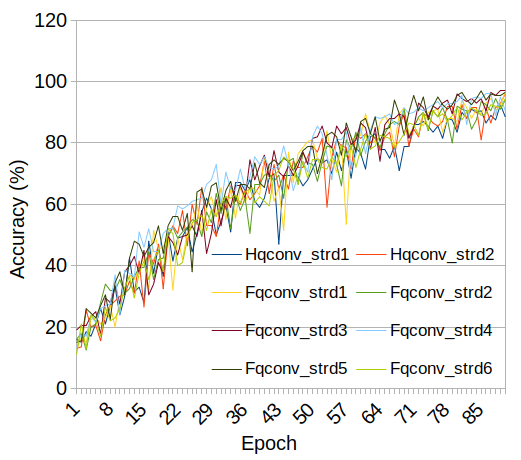}
		\caption{\centering Accuracy of HQconvs and FQconvs}
	\end{subfigure}
	\begin{subfigure}[b]{0.4\linewidth}
		\includegraphics[width=\linewidth]{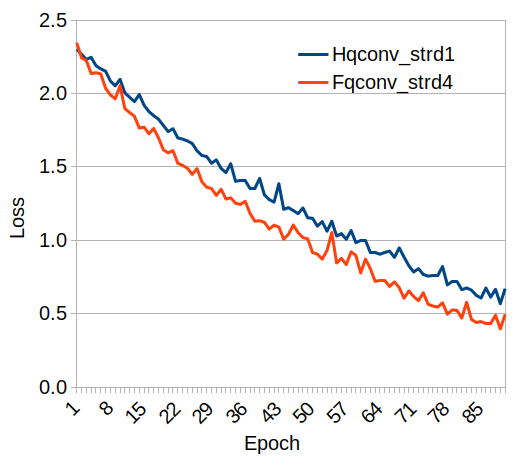}
		\caption{\centering Loss of HQconv (stride=1) and FQconv (stride=4)}
	\end{subfigure}
	\begin{subfigure}[b]{0.4\linewidth}
		\includegraphics[width=\linewidth]{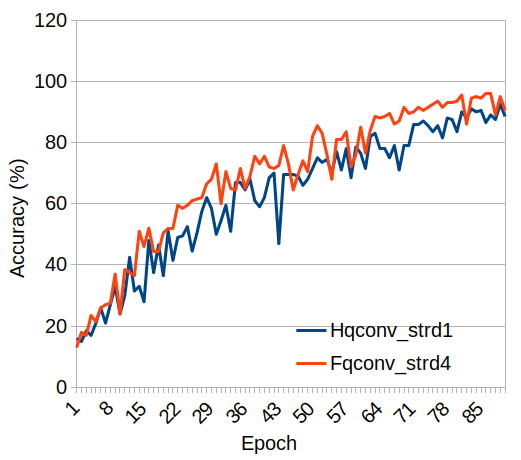}
		\caption{\centering Accuracy of HQconv (stride=1) and FQconv (stride=4)}
	\end{subfigure}
	\caption{Performances of quantum convolutional neural networks using HQconvs and FQconvs with different strides on CIFAR-10-small dataset (resolution $10 \times 10$). FQconv with stride set as $4$ outperforms most other Qconvs.}
	\label{fig:HQconv vs FQconv}
\end{figure}


\noindent \textbf{Experimental environment}   The experiments  in this paper are conducted on the Kunfeng Kungang\texttrademark \space hybrid classical-quantum cloud platform,  which supports large scale full-amplitude simulations of quantum circuits up to 42 qubits, as well as  classical DNN trainings on GPUs.
For the specific experiments in this paper,  we uses 4 Intel Xeon Processor (Skylake, IBRS) CPUs (128 logical cores) and 1,024GB of memory for quantum circuit simulations, and  Tesla V100-SXM2 GPU with 16 GB of memory for classical network training.


\begin{figure}[htpb]
	\centering
	\begin{subfigure}[b]{0.48\linewidth}
		\includegraphics[width=\linewidth]{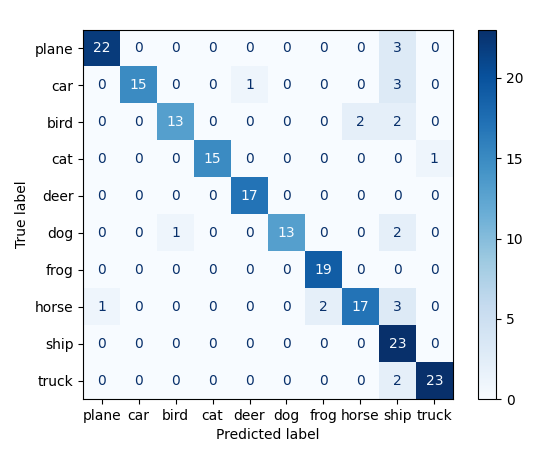}
		\caption{\centering HQconv (stride=1)}
	\end{subfigure}
	\begin{subfigure}[b]{0.48\linewidth}
		\includegraphics[width=\linewidth]{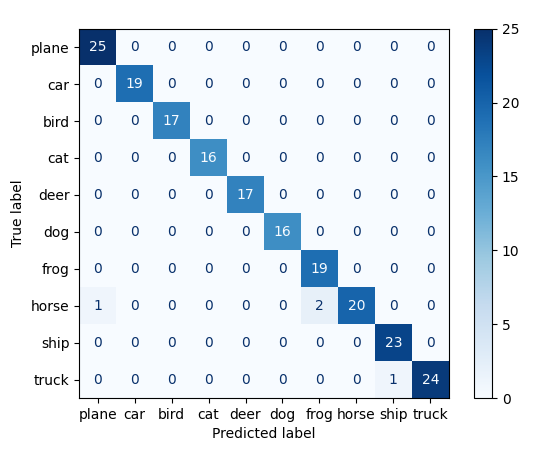}
		\caption{\centering FQconv (stride=4)}
	\end{subfigure}
	\caption{Confusion matrices of quantum convolutional neural networks use HQconv (stride=$1$) and FQconv (stride=$4$) on CIFAR-10-small dataset (resolution $10 \times 10$) after $90$ epochs. FQconv achieves higher predictive accuracy than HQconv.}
	\label{fig:ConfusionMtrx}
\end{figure}

\begin{figure}[htpb]
	\centering
	\begin{subfigure}[b]{0.48\linewidth}
		\includegraphics[width=\linewidth]{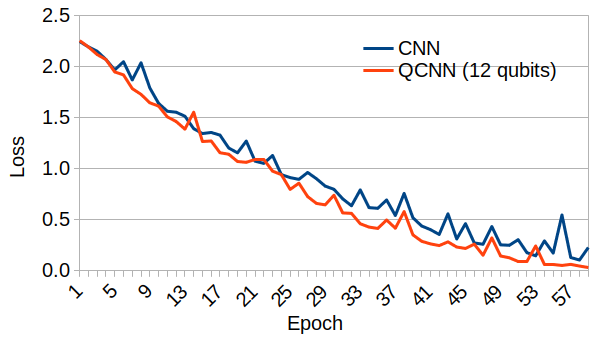}
		\caption{\centering Loss (CIFAR-10-small resolution $10 \times 10$) \newline}
	\end{subfigure}
	\begin{subfigure}[b]{0.48\linewidth}
		\includegraphics[width=\linewidth]{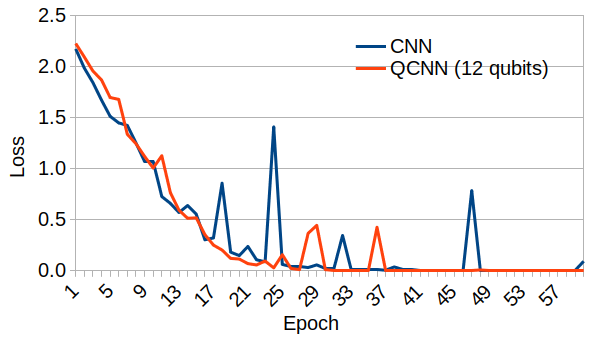}
		\caption{\centering Loss (CIFAR-10-small resolution $20 \times 20$)}
	\end{subfigure}
	\begin{subfigure}[b]{0.48\linewidth}
		\includegraphics[width=\linewidth]{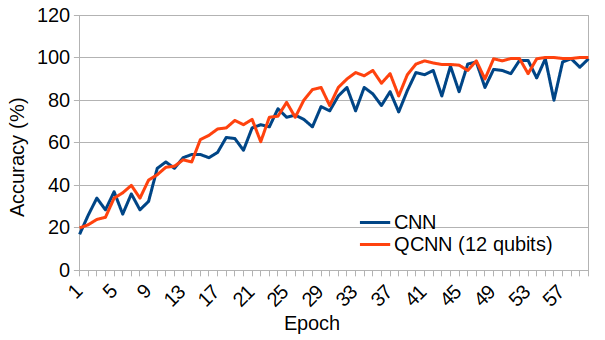}
		\caption{\centering Accuracy (CIFAR-10-small resolution $10 \times 10$)}
	\end{subfigure}
	\begin{subfigure}[b]{0.48\linewidth}
		\includegraphics[width=\linewidth]{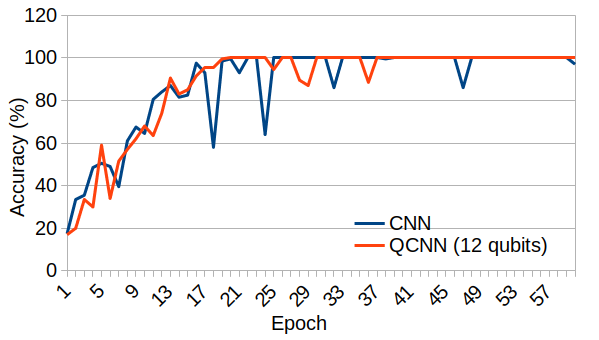}
		\caption{\centering Accuracy (CIFAR-10-small resolution $20 \times 20$)}
		
	\end{subfigure}
	\caption{Performance of quantum convolutional neural networks using HQconv together with classical convolutional layers on CIFAR-10-small dataset. (a, b) Losses of hybrid quantum-classical convolutional neural networks decrease in a smoother way than classical convolutional neural networks.}
	\label{fig:HQconv vs CNN}
\end{figure}

\begin{table}[]
	\centering
	\begin{tabular}{@{}cccc@{}}
		\toprule
		\thead{\textbf{Resolution}}      & \thead{\textbf{Curve}}       &
		\thead{ \textbf{Averaged} \\ \textbf{$\ell_1$-norm}}  & \thead{\textbf{Standard} \\
			\textbf{Deviation} \\ } \\ \midrule
		\multirow{4}{*}{10 x 10} & CNN\_loss               & 0.053          & 0.047                       \\
		& \textbf{QCNN\_loss}     & \textbf{0.035} & \textbf{0.030}              \\
		& CNN\_accuracy           & 3.179          & 2.484                       \\
		& \textbf{QCNN\_accuracy} & \textbf{2.005} & \textbf{1.607}              \\ \midrule
		\multirow{4}{*}{20 x 20} & CNN\_loss               & 0.078          & 0.135                       \\
		& \textbf{QCNN\_loss}     & \textbf{0.032} & \textbf{0.045}              \\
		& CNN\_accuracy           & 3.094          & 4.318                       \\
		& \textbf{QCNN\_accuracy} & \textbf{1.853} & \textbf{3.176}              \\ \midrule
	\end{tabular}
	\caption{Quantitative evaluation of smoothness for curves plotted in Fig.
		\ref{fig:HQconv vs CNN}. Comparing with curves from classical convolutional neural networks, all of loss and accuracy curves from the hybrid quantum-classical convolutional neural networks are closer to their corresponding baseline curves processed by Savitzky–Golay filter, which means the curves from QCNN are more smooth.}
	\label{tb:smoothness vs cnn}
\end{table}

\begin{figure}[htpb]
	\centering
	\begin{subfigure}[b]{0.48\linewidth}
	\includegraphics[width=\linewidth]{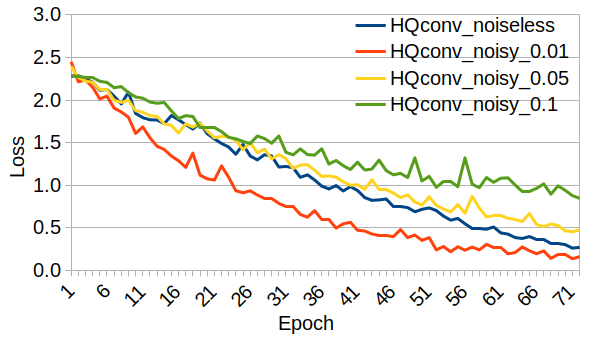}
	\caption{\centering Loss by HQconv w/wo Noise}
	\end{subfigure}
	\begin{subfigure}[b]{0.48\linewidth}
		\includegraphics[width=\linewidth]{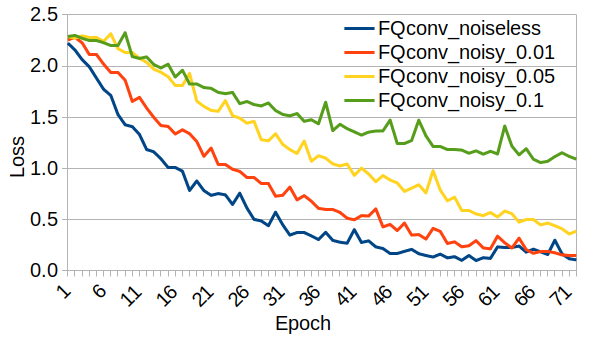}
		\caption{\centering Loss by FQconv w/wo Noise}
	\end{subfigure}
	\begin{subfigure}[b]{0.48\linewidth}
		\includegraphics[width=\linewidth]{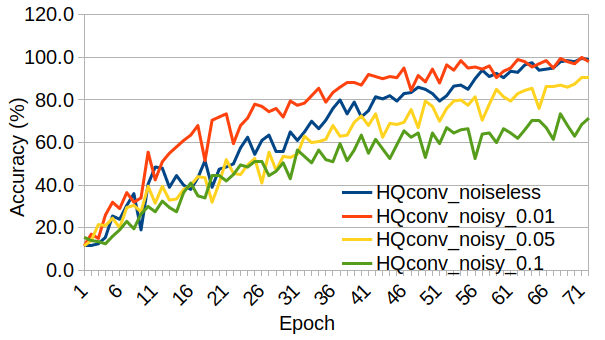}
		\caption{\centering Accuracy by HQconv w/wo Noise}
	\end{subfigure}
	\begin{subfigure}[b]{0.48\linewidth}
		\includegraphics[width=\linewidth]{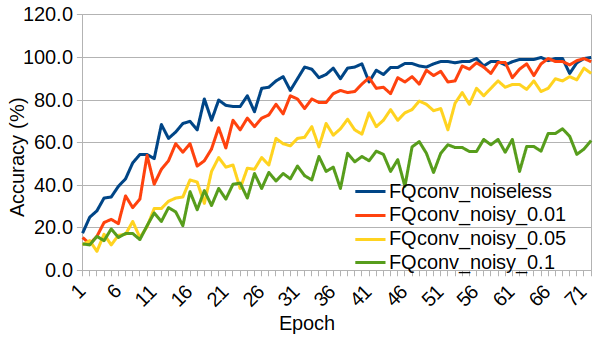}
		\caption{\centering Accuracy by FQconv w/wo Noise}
	\end{subfigure}
	\caption{Classification performances from quantum convolutional neural networks w/wo noise. Dataset used is CIFAR-10-small dataset, resolution is $10 \times 10$. HQconv ($stride=1$), FQconv ($stride=4$) are used in the experiments.}
	\label{fig:cifar10_10x10_12q_noisy}
\end{figure}

\subsection{Comparison among HQconv, FQconv, and classical CNNs}

As shown in Figure \ref{fig:HQconv vs FQconv}, both HQconvs and FQconvs achieve high classification accuracy on randomly-sampled CIFAR-10-small dataset, and FQconvs are generally better than HQconvs.
From the experimental results, the choice of stride has limited influence on HQconvs, but matters to FQconv w.r.t classification accuracy and loss.
Specifically, $strd = 4$ is the optimal stride value with the lowest loss curve for FQconvs, and the one with other stride values achieve close but slightly worse results.
In particular, FQconv outperforms other Qconvs when the stride is set up as $4$ or $5$, which indicates processing intra-channel information with controlled quantum gates is more important than processing inter-channel information.
From the confusion matrices exhibited in Figure \ref{fig:ConfusionMtrx}, FQconv with stride set to $4$ predicts labels more accurately than those of HQconv with stride set to $1$.

\begin{figure}[htpb]
	\centering
	\begin{subfigure}[b]{0.48\linewidth}
		\includegraphics[width=\linewidth]{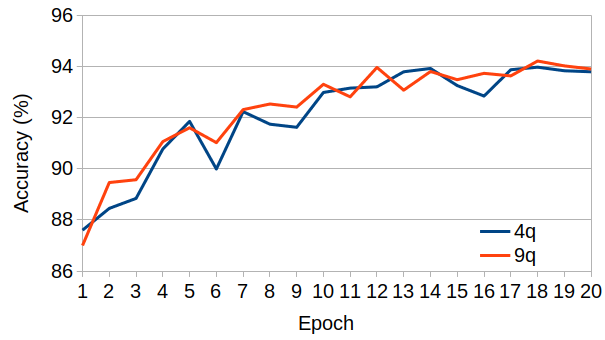}
		\caption{\centering Accuracy (MNIST resolution $20 \times 20$)}
	\end{subfigure}
	\begin{subfigure}[b]{0.48\linewidth}
		\includegraphics[width=\linewidth]{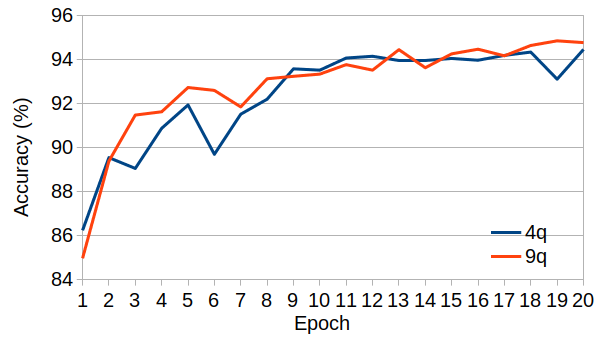}
		\caption{\centering Accuracy (MNIST resolution $28 \times 28$)}
	\end{subfigure}
	\begin{subfigure}[b]{0.48\linewidth}
		\includegraphics[width=\linewidth]{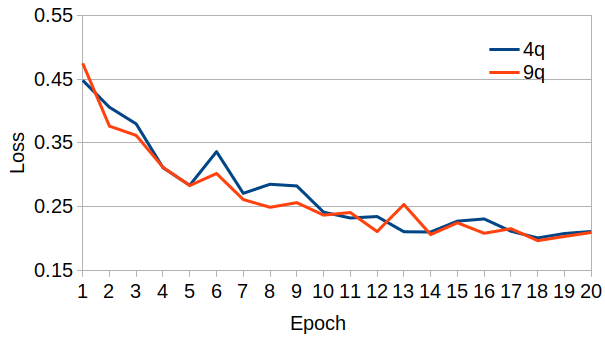}
		\caption{\centering Loss (MNIST resolution $20 \times 20$)}
	\end{subfigure}
	\begin{subfigure}[b]{0.48\linewidth}
		\includegraphics[width=\linewidth]{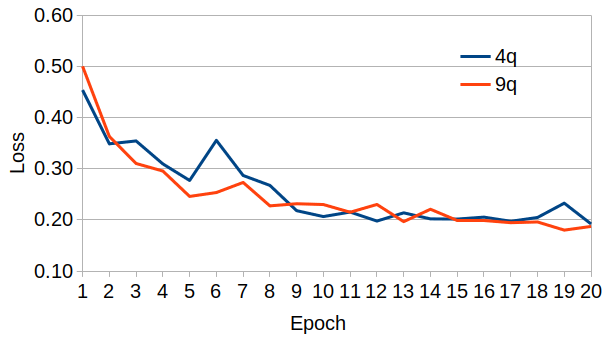}
		\caption{\centering Loss (MNIST resolution $28 \times 28$)}
	\end{subfigure}
	\caption{Comparison of classification performances from quantum convolutional neural networks with distinct kernel sizes on MNIST. Quantum circuit ansatz with $9$ qubits [kernel size $(3, 3)$] is slightly better than the one with $4$ qubits [kernel size $(2, 2)$].}
	\label{fig:mnist_4qvs9q}
\end{figure}

\begin{figure}[htpb]
	\centering
	\begin{subfigure}[b]{0.48\linewidth}
		\includegraphics[width=\linewidth]{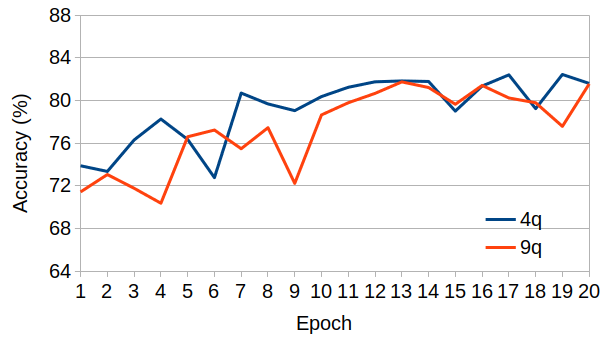}
		\caption{\centering Accuracy (fashion-MNIST resolution $20 \times 20$)}
	\end{subfigure}
	\begin{subfigure}[b]{0.48\linewidth}
		\includegraphics[width=\linewidth]{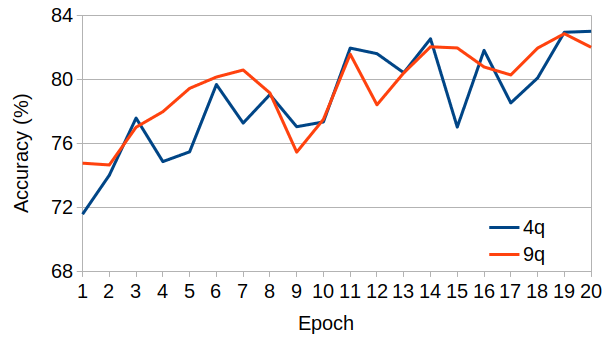}
		\caption{\centering Accuracy (fashion-MNIST resolution $28 \times 28$)}
	\end{subfigure}
	\begin{subfigure}[b]{0.48\linewidth}
		\includegraphics[width=\linewidth]{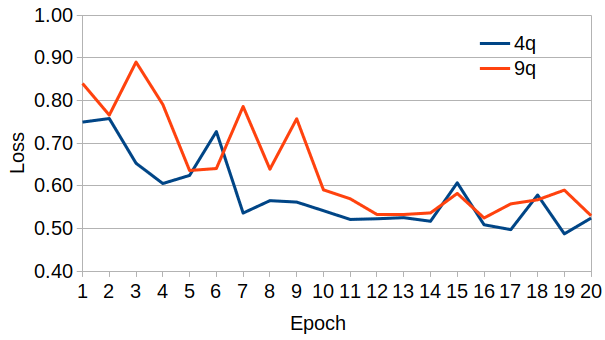}
		\caption{\centering Loss (fashion-MNIST resolution $20 \times 20$)}
	\end{subfigure}
	\begin{subfigure}[b]{0.48\linewidth}
		\includegraphics[width=\linewidth]{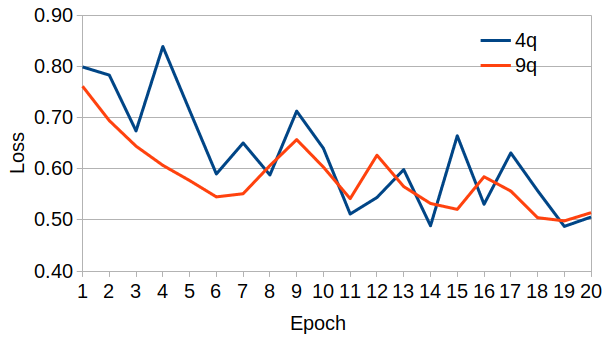}
		\caption{\centering Loss (fashion-MNIST resolution $28 \times 28$)}
	\end{subfigure}
	\caption{Comparison of classification performances from quantum convolutional neural networks with distinct kernel sizes on MNIST. Quantum circuit ansatz with $9$ qubits [kernel size $(3, 3)$] learns to classify images faster than the one with $4$ qubits [kernel size $(2, 2)$] when images are not compressed.}
	\label{fig:fmnist_4qvs9q}
\end{figure}

\begin{figure}[htpb]
	\centering
	\begin{subfigure}[b]{0.48\linewidth}
		\includegraphics[width=\linewidth]{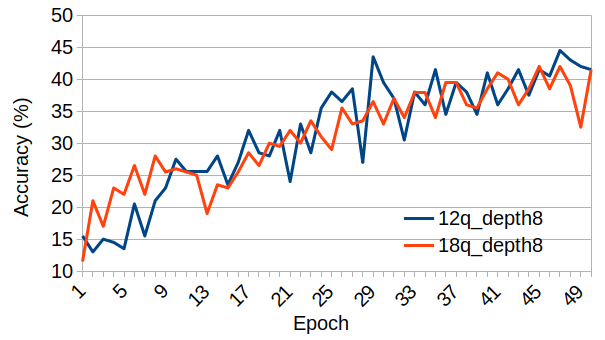}
		\caption{\centering Accuracy (CIFAR-10-small resolution $10 \times 10$)}
	\end{subfigure}
	\begin{subfigure}[b]{0.48\linewidth}
		\includegraphics[width=\linewidth]{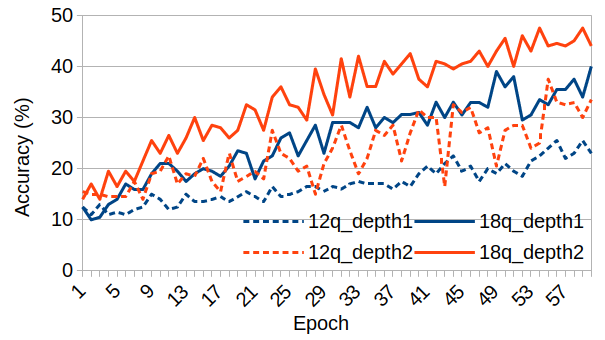}
		\caption{\centering Accuracy (CIFAR-10-small resolution $20 \times 20$)}
	\end{subfigure}
	\begin{subfigure}[b]{0.48\linewidth}
		\includegraphics[width=\linewidth]{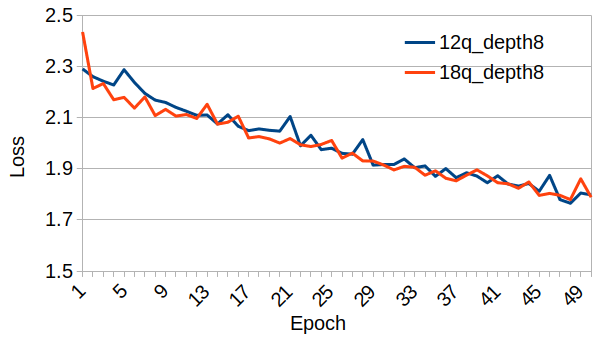}
		\caption{\centering Loss (CIFAR-10-small resolution $10 \times 10$) \newline}
	\end{subfigure}
	\begin{subfigure}[b]{0.48\linewidth}
		\includegraphics[width=\linewidth]{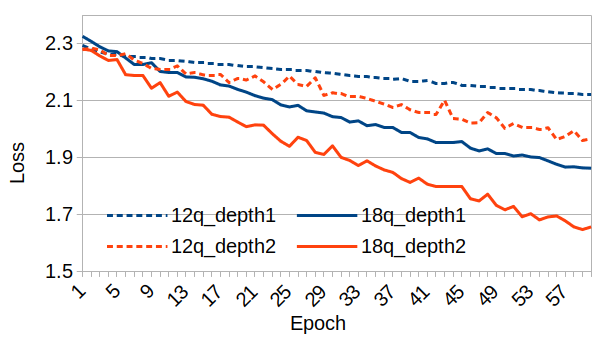}
		\caption{\centering Loss (CIFAR-10-small resolution $20 \times 20$)}
	\end{subfigure}
	\caption{Comparison of classification performances from quantum convolutional neural networks with distinct kernel sizes on the CIFAR-10-small dataset. HQconvs with stride 1 are used in the experiments.}
	\label{fig:CIFAR-10_10x10_12qvs18q}
\end{figure}

As shown in Figure \ref{fig:HQconv vs CNN}, our hybrid model trained with size $10 \times 10$ images on the CIFAR-10-small dataset achieves a higher accuracy than the corresponding classical convolutional neural network. Comparing with classical convolutional neural networks, the losses of the hybrid models with HQconv decrease more smoothly and the accuracies from the hybrid models are more stable on the CIFAR-10-small dataset.
To see this quantitatively, for each curve in raw data, we use Savitzky–Golay filter to smooth it to generate a baseline curve. Then we compare the raw data curve with the baseline curve by calculating the averaged $\ell^1$-norm and the standard deviation of the gap between them.
The average is over the number of data points.
A smaller averaged $\ell^1$-norm and standard deviation indicate the curve is more smooth, the results of which are in Table \ref{tb:smoothness vs cnn}.


As shown in Figure \ref{fig:cifar10_10x10_12q_noisy}, when the noise level is under 0.05, the hybrid models achieve similar predictive accuracy compared with the hybrid models without noise, which is because small noise has a limited impact on the training process and the predictive accuracy of the quantum circuits. We observe that for circuits with noise greater than 0.1, the existence of noise not only slows the loss convergence but also decreases the predictive accuracy. 
As shown in Figure \ref{fig:cifar10_10x10_12q_noisy} (a) and (c), the loss for HQconv with a noise level set as 0.01 outperforms HQconv without noise, and the accuracies are almost the same for both cases. The reason of loss function for noise level 0.01 performs better than HQconv without noise is that the landscape of the loss function may be full of local minimums, many paths lead to local minimums, and sometimes a good descent direction overcomes the impact from the small noise. We repeat the experiments a few times and observe that the loss for HQconv with noise level 0.01 underperforms HQconv without noise sometimes.

By including extra convolutional layers like the architecture shown in Figure \ref{fig:NQCNN archit}, classification accuracy could be further improved, duration of the training procedure is also shortened.

\subsection{Classification Performances with distinct Kernel Sizes}

Kernel size determines how fine-grained the features are extracted from the input. Smaller kernels keep more detailed information than larger ones do.
On the other hand, too many details could fluctuate the performance of a image classifier.

Through  experiments, we observe that larger kernel size is more suitable for higher resolution images.
From results on datasets CIFAR-10-small, MNIST and fashion-MNIST, larger quantum ansaetze achieve better classification accuracies in general cases.
As shown in Figure \ref{fig:mnist_4qvs9q} and Figure \ref{fig:fmnist_4qvs9q}, quantum convolutional neural networks with $3 \times 3$ kernel converge at a faster rate at the beginning stage of the training than quantum convolutional neural networks with $2 \times 2$ kernels do, especially when images are not compressed.
Even for the CIFAR-10-small dataset (Figure \ref{fig:CIFAR-10_10x10_12qvs18q} (a)), quantum convolutional neural networks with a larger kernel is slightly better than quantum convolutional neural networks with a small kernel, when images are compressed to size 10 $\times$ 10.

The expressive capability of a quantum ansatz determines its ability to learn.
To further understand the expressive capability gap between large quantum ansatz (18 qubits, correspond to kernel size 2 $\times$ 3) and small quantum ansatz (12 qubits, correspond to kernel size 2 $\times$ 2), we use only one or two filters in the quantum convolutional layers.
By using a limited number of filters in the quantum convolutional layer, the extra expressiveness can be eliminated.
As shown in Figure \ref{fig:CIFAR-10_10x10_12qvs18q} (b, d), classification accuracy from an 18-qubit single-depth quantum convolutional neural network stays above 20$\%$ since $20^{th}$ epoch. However, for a 12-qubit single-depth quantum convolutional neural network, the same classification accuracy is achieved after $40$ epochs of training. Hence the quantum convolutional neural networks using larger kernels not only converge faster but also achieve higher accuracy than the ones with smaller kernels. This also indicates that quantum convolutional neural networks with larger kernels have better expressive capability than networks with smaller kernels.

\section{Conclusions and Future Work}

In this work, we introduce two types of quantum circuit ansaetze, HQconvs and FQconvs, for convolution operation applied on RGB images, which brings quantum computers into a colorful world. We investigate the order to process inter-channel and intra-channel information and observe that encoding intra-channel information is critical to improve the performance of the hybrid quantum-classical convolutional neural networks. For FQconvs, there exists an ideal stride value to optimize the networks. By gaining a deep understanding of different roles played by inter- and intra-channel information on network performance, it may inspire more effective architectures in quantum circuit ansatz design in future work. The proposed quantum circuit ansaetze show competitive predictive accuracies compared to classical CNNs.  We also find that
the larger quantum ansatz outperforms the smaller quantum ansatz on higher resolution image datasets. Our finding could provide some guidelines to choose kernel size in a quantum convolutional layer.

The proposed ansaetze require reasonable resources even with a relatively high
resolution of input images, which makes our system easier to be simulated with
classical computers or implemented with real quantum hardware. As IBM unveils breakthrough 127-Qubit quantum processor with gate error touching 0.001 \cite{ibm_error}, the proposed or similar ansatz circuits could achieve lower enough noise level and relatively good predictive accuracy on IBM or other near-term QPUs. This makes the ansaetze practical to real problems in the NISQ era. 

 For further work, it is very beneficial to reveal the correlation among the number of quantum gate operations, the circuit depth, and the kernel size of the QCNN ansatz circuit. With such a relationship, the quantitative estimation of the error of an ansatz circuit on specific hardware platforms can be performed. Combining the information, following the proposed methods of our current work, the systematic estimation of the QCNN performance on the real hardware can be achieved. This will give researchers a quantitative way to estimate a QPU’s capability of running QCNN circuits or other QNN circuits, which will be of great interest. This further investigation is beyond the scope of our current work, and we plan to tackle it as well as to apply the proposed ansaetze on NISQ quantum devices in our next-stage research.

%
%
%
%
%
%
%
%
%
%

\begin{acknowledgements}
We acknowledge Dr. Yi Zhou for insightful discussions.
The work is partially supported by Pudong New Area Science and Technology Development Fund, No. PKX2020-R17.

\end{acknowledgements}

\section*{Declarations}

%
\noindent {\bf Conflict of interest}

The authors declare that they have no conflict of interest.

\noindent {\bf Data availability}

Datasets used are available on websites \url{https://www.tensorflow.org/datasets/catalog/overview#image_classification}.

\noindent {\bf Code availability}

The code for this project is available in the GitLab repository found in
\url{https://gitlab.com/kunfeng}.

%
%

\bibliographystyle{unsrt2authabbrvpp}

\bibliography{refer}   


\end{document}